\documentclass[aps,prx,twocolumn,superscriptaddress,nofootinbib,showkeys]{revtex4-2}

\usepackage{graphicx}
\usepackage{dcolumn}
\usepackage{bm}
\usepackage[utf8]{inputenc}
\usepackage{textcomp}
\usepackage{amsmath}
\usepackage[mathlines]{lineno}
\usepackage{siunitx}
\usepackage{hyperref}
\usepackage{placeins}
\hypersetup{
    colorlinks=true,
    citecolor=blue,
    urlcolor=blue,
    linkcolor=black
}

\AtBeginDocument{\nocite{achemso-control}}

\begin{document}
\title{Waveguide-coupled light emitting devices made of 2D materials}

\author{Ronja Khelifa}
\thanks{These two authors contributed equally}
\affiliation{Photonics Laboratory, ETH Zürich, 8093 Zürich, Switzerland }
\author{Shengyu Shan}
\thanks{These two authors contributed equally}
\affiliation{Photonics Laboratory, ETH Zürich, 8093 Zürich, Switzerland }
\author{Takashi Taniguchi}
\affiliation{International Center for Materials Nanoarchitectonics, National Institute for Materials Science, 1-1 Namiki, Tsukuba 305-0044, Japan}
\author{Kenji Watanabe}
\affiliation{Research Center for Functional Materials, National Institute for Materials Science, 1-1 Namiki, Tsukuba 305-0044, Japan}
\author{Lukas Novotny}
\affiliation{Photonics Laboratory, ETH Zürich, 8093 Zürich, Switzerland }
\date{\today}

\begin{abstract}

Optical information processing using photonic integrated circuits is a key goal in the field of nanophotonics. Extensive research efforts have led to remarkable progress in integrating active and passive device functionalities within one single photonic circuit. Still, to date, one of the central components - i.e. light sources - remain a challenge to be integrated. Here, we demonstrate waveguide-coupled electrically driven light emitters in an on-chip photonics platform based on 2D materials. We combine light-emitting devices (LEDs), based on exciton recombination in transition metal dichalcogenides (TMDs), with hexagonal boron nitride (h-BN) photonic waveguides in a single van der Waals (vdW) heterostructure. Waveguide-coupled light emission is achieved by sandwiching the LED between two h-BN slabs and patterning the complete vdW stack into a photonic structure. Our demonstration of on-chip light generation and waveguiding is a key component for future integrated vdW optoelectronics.
\end{abstract}


\maketitle

With the rapid advances in the field of optical information processing, the demand for the development of integrated photonic circuits is constantly increasing. Along with photodetectors and optical modulators, a fundamental building block are integrated light sources, which have, however, remained a challenge and led to extensive research \cite{Dong2014,Komljenovic2018,Elshaari2020,Liu2021,Han2022}. 
One suitable material class that has shown great potential for integrated optoelectronic devices are two-dimensional (2D) materials \cite{Xia2014,Koppens2014,Bie2017,Flory2020,Thakar2019,An2022,Yang2022}.
Among others, one benefit is the vertical assembly of dissimilar 2D materials to form van der Waals (vdW) heterostructures, that can be placed on almost any substrate without constraints of lattice matching \cite{Geim2013}. The resulting possibility of band gap engineering with atomic precision has opened up a new level of control for the design of light emitting quantum wells \cite{Withers2015,Withers2015a,Liu2017}, tunnel junctions \cite{Kuzmina2021} and other optoelectronic devices \cite{Ross2017,Jauregui2019,Mak2016,Mueller2018}.\par

\begin{figure}[!t]
\centering
\includegraphics[width=1\columnwidth]{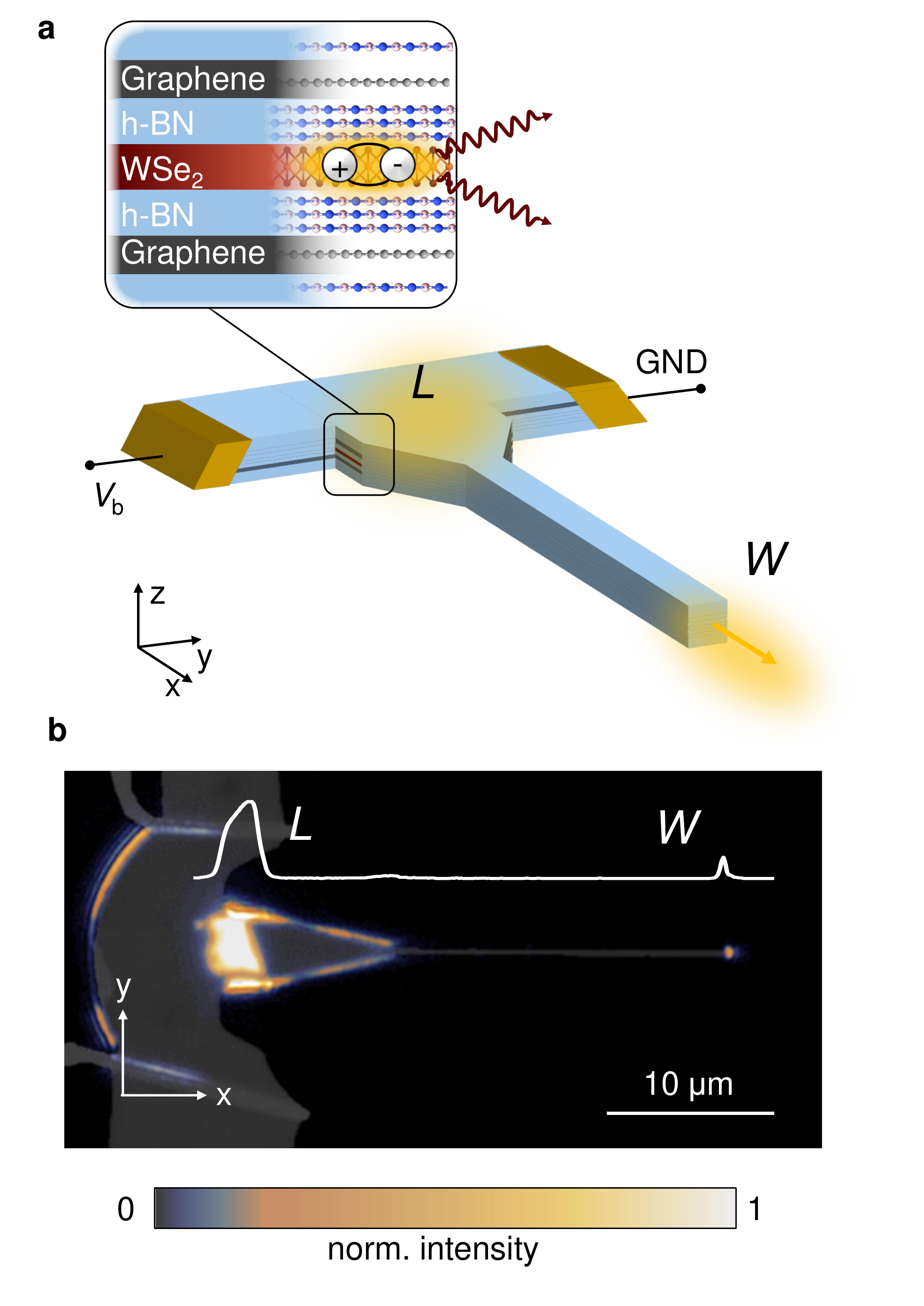}
\caption{(a) Illustration of the waveguide-coupled LED. A WSe\textsubscript{2}-based tunneling device is integrated near the vertical center of an h-BN waveguide structure. (b) Real space image of the LED's emission ($V_\text{b}=\SI{1.7}{\volt}$), overlaid with the device shape. The inset shows the intensity profile along the central section of the waveguide-coupled device (x-direction). The LED region is labeled \textit{L} and the waveguide end is labeled \textit{W}.} \label{fig:Schematic}
\end{figure}

\begin{figure*}[!ht] 
\centering
\includegraphics[width=1\textwidth]{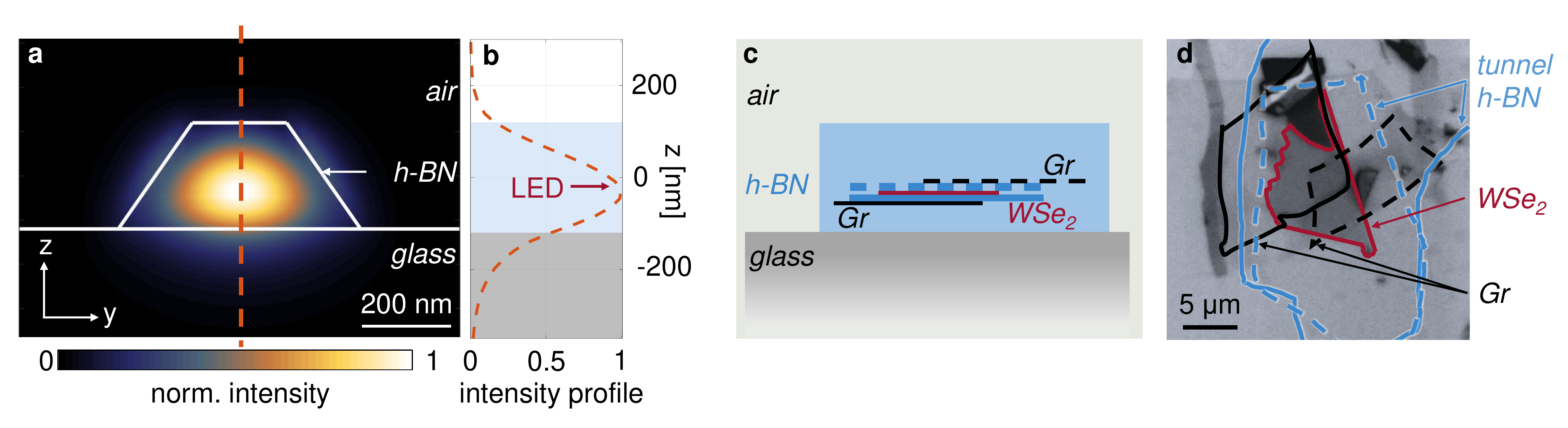}
\caption{(a) Intensity distribution in an h-BN waveguide placed on top of a glass substrate. (b) Intensity profile along the z-direction (orange dashed line in (a)), whereby $z=\SI{0}{\nano \meter}$, is the vertical center of the waveguide. The peak maximum is shifted towards the glass substrate, due to its higher refractive index. The LED is placed near the highest field intensity, as indicated by the red arrow. (c) Schematic cross-section and (d) reflection map of the vdW heterostructure after stacking: top h-BN ($130-\SI{140}{\nano \meter}$) -- Gr (1 layer) -- h-BN (4$\pm$1 layers) -- WSe\textsubscript{2} (1 layer) -- h-BN (4$\pm$1 layers) -- Gr (1 layer) -- bottom h-BN ($\sim$\SI{105}{\nano \meter}).}  \label{fig:stack}
\end{figure*}\par

The co-integration of light emitting devices (LEDs) with on-chip photonics has been demonstrated \cite{Bie2017,Liu2017,Yang2022,GonzalezMarin2022}, for example by placing 2D materials on top of silicon-based waveguides to couple their emission via evanescent fields \cite{Bie2017}. However, placing the active material inside the light confining dielectric, where the field intensity is higher, improves the mode overlap between emitter and waveguide mode \cite{Ye2015, Ren2018, Khelifa2020}, and thereby the coupling efficiency \cite{Li2021a}. For optically excited systems, this has been achieved by sandwiching monolayer transition metal dichalcogenides (TMDs) as active materials inside a free-standing microdisk made of Si\textsubscript{3}N\textsubscript{4}/hydrogen silsesquioxane (HSQ) \cite{Ye2015}. A similar approach was adopted for photonic devices entirely made of 2D materials, namely by sandwiching TMDs between thick slabs of hexagonal boron nitride (h-BN) \cite{Ren2018, Ren2019, Khelifa2020}.
The h-BN platform has evolved as an attractive candidate for on-chip photonics \cite{Kim2018,Froch2018a,Lassaline2021} for which moreover the monolithic integration of quantum emitters has been demonstrated \cite{Li2021a}.
However, all of these structures rely on optical excitation with external lasers, which prevents full-scale integration.

Here we demonstrate the integration of electrically-driven LEDs with h-BN photonic waveguides. 
As illustrated in Fig.~\ref{fig:Schematic}a, we focus on a photonic platform that is solely based on 2D materials. We integrate TMD-based LEDs inside h-BN waveguide structures, such that a high mode overlap is guaranteed. As a result, light emitted by the LED is coupled into the h-BN waveguide. The radiation emitted from this electrically-driven device is shown in Fig.~\ref{fig:Schematic}b. The LED is confined to the region labeled \textit{L} and the spot at the far right (labeled \textit{W}) corresponds to out-coupled light at the end of the h-BN waveguide.
Our fully integrated device combines two major properties: 1) waveguide-coupled emission from an electrically driven light source and 2) enhanced optical mode overlap between emitter and guided mode. In the following, we first elaborate on the integration of the LED near the highest field intensities for enhanced mode overlap. Next, we describe the properties of an unpatterned LED (with no waveguide coupling), and then compare it with the same device after patterning.\par

 
Our LED (inset in Fig.~\ref{fig:Schematic}a) is sandwiched between thick layers of h-BN. These h-BN layers define the vertical height of the photonic waveguide. As shown in Figs.~\ref{fig:stack}a, b the intensity of the waveguide mode reaches its maximum close to the vertical center of the waveguide. By placing the LED at the indicated position inside the dielectric, we increase the mode overlap between emitter and waveguide mode to achieve more efficient coupling.  
To fabricate the device illustrated in Fig.~\ref{fig:Schematic}a, flakes of graphene (Gr), WSe\textsubscript{2} and h-BN were mechanically exfoliated and then stacked using a polymer-based stacking technique \cite{Wang2013, Zomer2014}. Optical and atomic force microscopy were used to identify the desired thicknesses of the flakes. Figures~\ref{fig:stack}c and d, show a schematic and a reflection map of the vdW structure.
The thicknesses of the top and bottom encapsulating h-BN flakes were chosen such that the LED is located near the region of the highest optical field of the resulting h-BN waveguide mode. As visualized in Fig.~\ref{fig:stack}b, the field maximum is slightly misplaced from the vertical center of the waveguide ($z=\SI{0}{\nano \meter}$). The reason is that the refractive indices of the media outside of the encapsulating h-BN flakes are different, that is, the waveguide mode leaks more strongly into the glass substrate at the bottom than into the air at the top. Consequently, the bottom h-BN thickness ($\sim$\SI{105}{\nano \meter}) is chosen to be lower than the top h-BN thickness ($130-\SI{140}{\nano \meter}$).\par


\begin{figure}[!ht] 
\centering
\includegraphics[width=1\columnwidth]{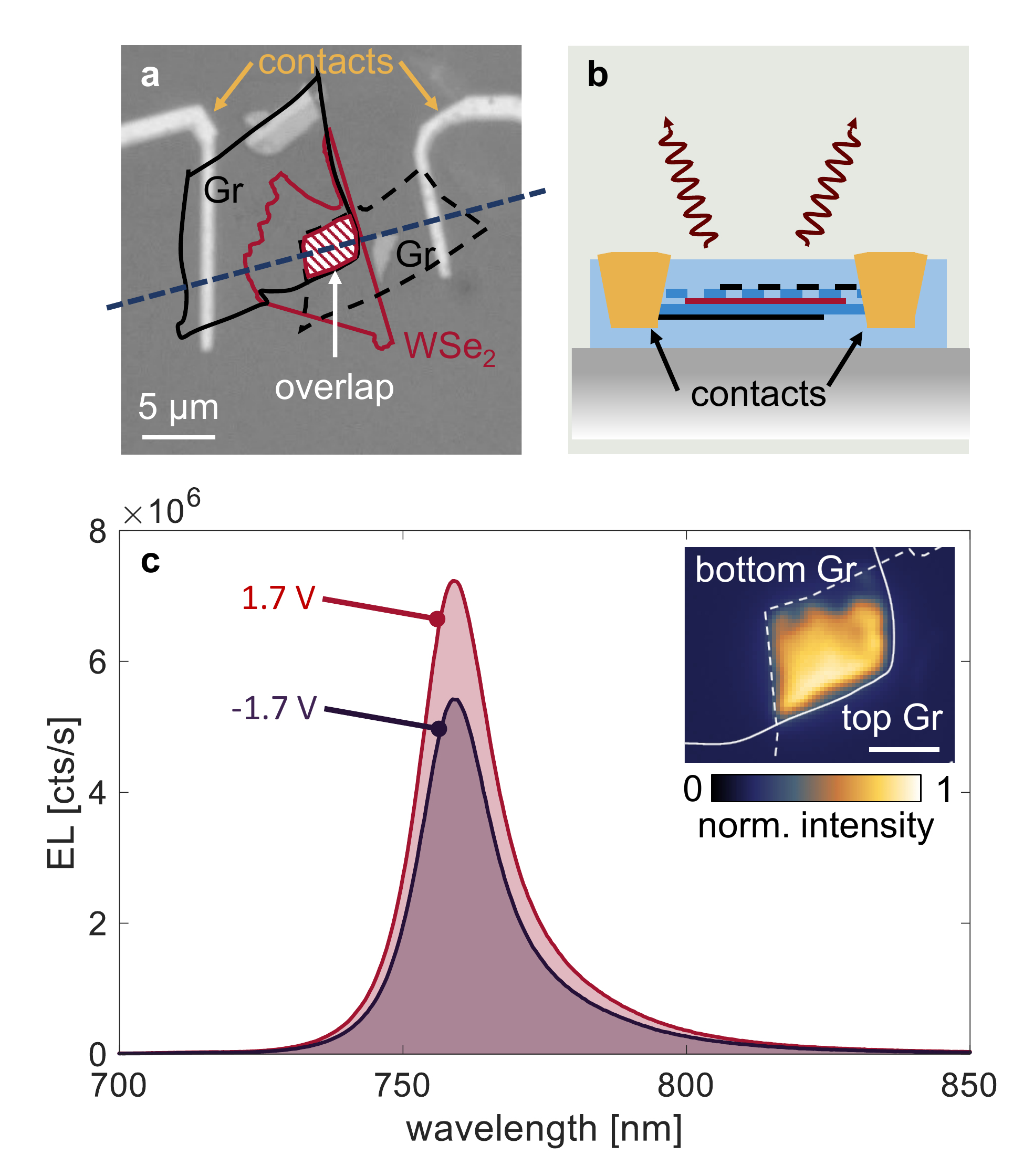}
\caption{(a) Optical image of the contacted LED after fabrication of the Gr edge contacts. The red-white striped area indicates the overlap region of all flakes. (b) Schematic cross-section of the LED, along the dashed blue line in (a). (c) EL spectra of the unpatterned device for a bias of \SI{1.7}{\volt} and \SI{-1.7}{\volt}. Inset: emission of the device for an applied bias of \SI{1.7}{\volt}. The emission area corresponds to the overlap region indicated in (a). The scale
bar is \SI{2}{\micro\meter}.}  \label{fig:unpatterned_LED}
\end{figure}\par

The device illustrated in Fig.~\ref{fig:Schematic}a entails the following fabrication steps: 1) stacking of the vdW heterostructure, 2) electrical contacting, and 3) patterning of the vdW device. For comparison, we also measured the unpatterned devices, i.e. devices before step 3.
As shown by the inset of Fig.~\ref{fig:Schematic}a, the LEDs consist of a monolayer WSe$_2$ sandwiched between two h-BN tunneling barriers (4$\pm$1 layers each) and two monolayer Gr electrodes \cite{Withers2015,Withers2015a, Liu2017}. The two Gr electrodes on top and bottom enable the vertical injection of charge carriers via tunneling through the thin h-BN barriers. The formation of excitons inside the WSe$_2$ monolayer and their subsequent radiative recombination leads to light emission from the overlap region \cite{Withers2015,Withers2015a}.\par
The embedded Gr electrodes are electrically contacted by edge contacts (Cr/Au) \cite{Wang2013,Overweg2018}, as illustrated in Figs.~\ref{fig:unpatterned_LED}a and b. Using the transfer length method (TLM), we estimate the contact resistance to be in the order of $200-\SI{300}{\ohm \micro \meter}$ (details in the Supporting Information).
These contacts are used to apply a bias voltage $V_\text{b}$. For $V_\text{b}=\SI{1.7}{\volt}$, electroluminescence (EL) is observed from the entire overlap region and is visualized in the real space image in the inset of Fig.~\ref{fig:unpatterned_LED}c. We attribute small variations in the emission region to inhomogeneities and bubbles between adjacent flakes. EL emission at voltages lower than the electronic band gap has already been reported for similar devices in Refs. \cite{Binder2017, Withers2015a}.
Figure~\ref{fig:unpatterned_LED}c shows representative EL spectra, which are symmetrical for both bias polarities, and only slightly differ in intensity. The emission is dominated by the neutral exciton, as also observable in photoluminescence (PL) measurements (see Fig.~S1a in the Supporting Information) \cite{Withers2015a,Karni2019}. Therefore, the electrically driven emission from the device can be attributed to the radiative recombination of excitons in the monolayer WSe$_2$.
The maximum external quantum efficiency of the device at a bias of $V_\text{b}=\SI{1.7}{\volt}$ is around 4~\%, which is in the same order of magnitude as the values previously reported for tunneling LEDs with monolayer WSe$_2$ at room temperature \cite{Withers2015a}.\par

\begin{figure}[!h]
\includegraphics[width=1\columnwidth]{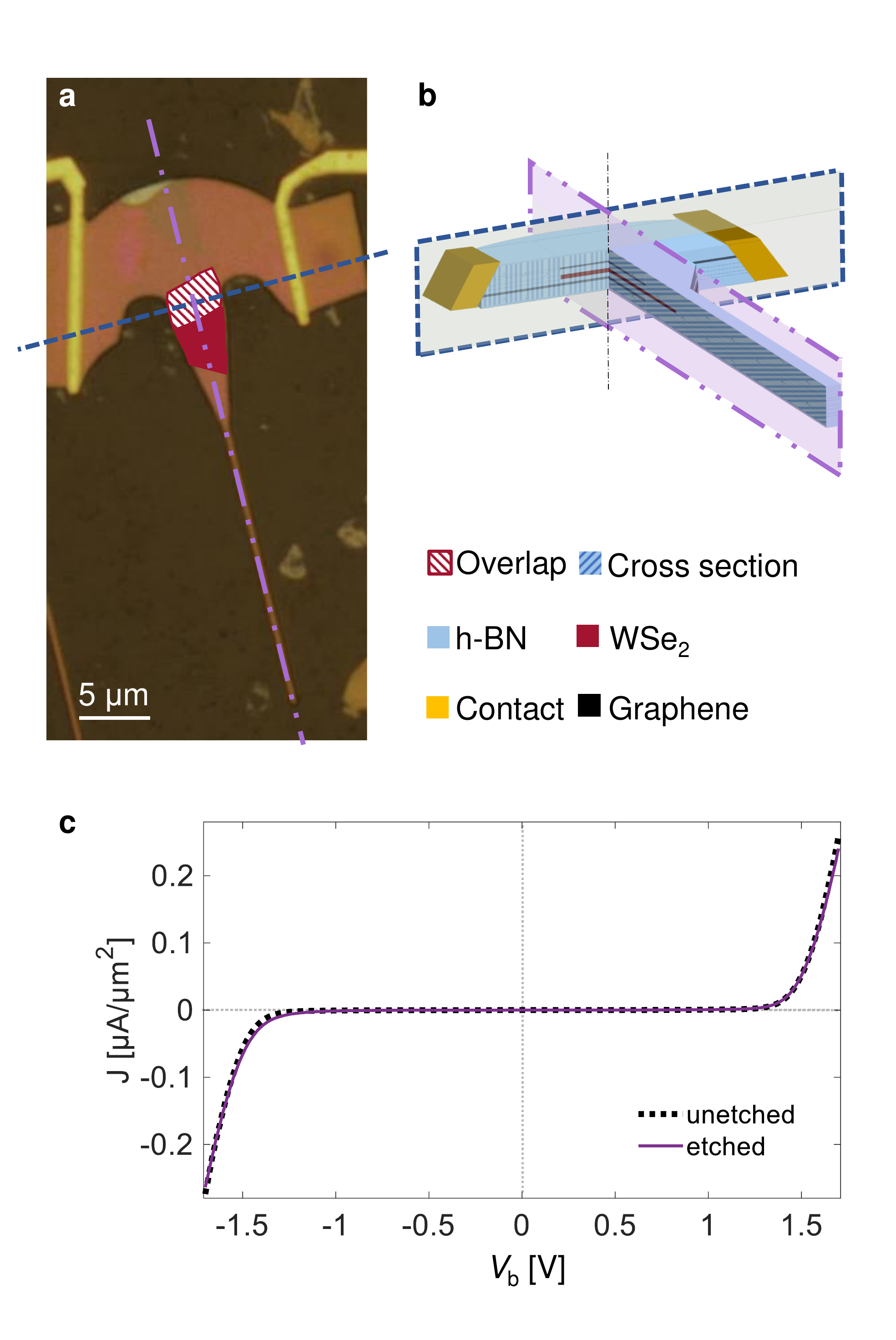}
\caption{(a) Optical microscope image of the final device after patterning the vdW stack. (b) Three-dimensional schematic of the final device. The blue and purple boxes indicate cross-sections in two directions, as indicated in (a). The photonic structure is aligned such, that the overlap region of the LED (red-white striped region in (a)) is centered and positioned in front of the waveguide taper. (c) $J$-$V$ characteristics of the final device after patterning (purple solid line) and before patterning (black dashed line).}  \label{fig:patterned_LED}
\end{figure}\par

\begin{figure*}[!t]
\includegraphics[width=1\textwidth]{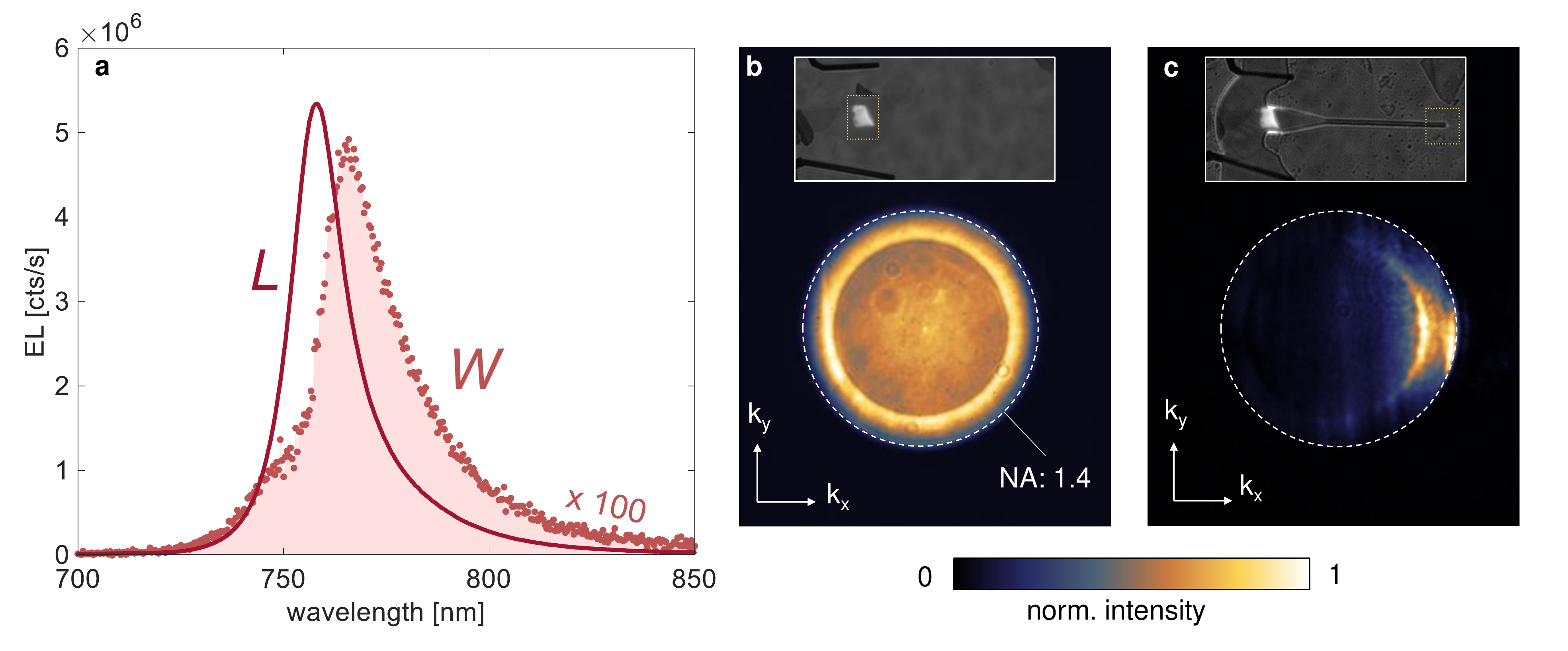}
\caption{ (a) Spectra of the waveguide-coupled emission (\textit{W}, red dotted line) and direct emission (\textit{L}, red solid line). (b) Fourier space image (radiation pattern) of the emission of the unpatterned LED and (c) of the patterned device at the end of the waveguide (region \textit{W}).}  \label{fig:Spectra_and_FourierSpace}
\end{figure*}


In a next step, the vdW stack containing the LED (inset in Fig.~\ref{fig:Schematic}a) was patterned into a photonic structure using electron beam lithography and reactive ion etching \cite{Khelifa2020}. This step utilizes the real space EL map of the unpatterned LED (inset of Fig.~\ref{fig:unpatterned_LED}c) as a reference for the device layout, which is favorable compared to fabrication methods that rely on the accurate transfer of vdW stacks on prefabricated photonic structures \cite{Bie2017, Flory2020}.
To increase light coupling we linearly taper the width of the device at the base of the h-BN waveguide (region \textit{L} in Fig.~\ref{fig:Schematic}) from \SI{3.8}{\micro \meter} to \SI{550}{\nano \meter}.
A microscope image and schematic of the final device after reactive ion etching are shown in Figs.~\ref{fig:patterned_LED}a and b. Details on the fabrication process are discussed in the Supporting Information.\par
After applying a bias voltage to the patterned device, we observe light emission from the end of the waveguide (region \textit{W} in Fig.~\ref{fig:Schematic}).
At the waveguide end the light is emitted into the glass substrate at high angles, which requires collection with an oil-immersion objective ($\text{NA}=1.4$).\par
Figure~\ref{fig:patterned_LED}c shows the current density ($J$) characteristics as a function of voltage ($V$) for both, the patterned (purple solid line) and unpatterned (black dashed line) device. The two $J$-$V$ curves are nearly identical, indicating that the etching process does not affect the device functionality.\par
The angular distribution of the light emission (radiation pattern) is measured by Fourier space imaging (Figs.~\ref{fig:Spectra_and_FourierSpace}b, c). For the waveguide-coupled EL of the patterned device, we observe light emission into large angles and with high directivity (Fig.~\ref{fig:Spectra_and_FourierSpace}c). This emission pattern agrees with the out-coupling of an in-plane propagating waveguide mode. In contrast, Fig.~\ref{fig:Spectra_and_FourierSpace}b shows the Fourier space image from the unpatterned device. Here, the EL emission pattern agrees with the emission pattern resulting from in-plane oriented emitters. Similar patterns are also observed for PL from monolayer WSe\textsubscript{2} (see Fig.~S1b in the Supporting Information) and agree with PL studies on TMD monolayers \cite{Schuller2013}.\par
Spatially resolved spectral measurements allow us to differentiate between free-space EL from the LED region (region \textit{L} in Fig.~\ref{fig:Schematic}) and the waveguide-coupled EL from the end facet of the waveguide (region \textit{W} in Fig.~\ref{fig:Schematic}). 
Spectra from both regions are shown in Fig.~\ref{fig:Spectra_and_FourierSpace}a.
Both spectra are peaked around the WSe\textsubscript{2} exciton wavelength.
However, the peak from region \textit{L} is red-shifted with respect to the peak from region \textit{W}. 
There are different possible explanations for this shift, which we discuss in the following. Using finite-difference time-domain (FDTD) simulations (Lumerical Inc.) we verified that the effect of waveguide and tapering has a negligible contribution on the spectrum (in this spectral window). Another factor is the spectral dependence of the out-coupling and collection efficiency of the emission at the waveguide end. Simulations show that the collection of the out-coupled waveguide mode through the glass substrate changes the spectrum to a small extent, but cannot be solely responsible for the shift.
In addition, the fabricated waveguides have some sidewall roughness, which leads to scattering losses that can impact the measured spectrum. Moreover, absorption of light by WSe\textsubscript{2} \cite{Kozawa2014} can affect the spectrum as well. This is because part of the WSe\textsubscript{2} is sticking out (red area in Fig.~\ref{fig:patterned_LED}a) of the overlap region (red-white striped area), penetrating into the tapered region where the light is coupled to. 
Such reabsorption losses can be reduced by an improved alignment of the individual layers of the stack.\par
The measurements yield an overall coupling efficiency of $\SI{0.3}{\percent}$. This value refers to the integrated spectra (see Supporting Information for more details). 
Our FDTD simulations indicate an estimated value of $\sim$\SI{1.2}{\percent} for the coupling efficiency (see Supporting Information), slightly higher than the measured value. 
The coupling efficiency can be improved by optimizing the lateral design of the patterned structure \cite{Fu2014,Andrade2019}. This, however, is above the scope of this work.\par

In conclusion, we demonstrate waveguide-coupled EL from an integrated LED in an all-2D material-based photonics platform. 
Our waveguide-coupled LEDs are fabricated by sandwiching a light emitting vdW heterostructure in between thick h-BN layers and subsequent patterning. To improve the coupling efficiency, we increase the mode overlap by placing the LED near the vertical center of the h-BN.\par 
Our device represents a versatile building block that can be further developed by integrating other vdW-based optoelectronic devices \cite{Kuzmina2021,Ross2017,Jauregui2019,Flory2020,Bie2017}. For example, combined with waveguide-coupled photodetectors \cite{Flory2020} our LED constitutes an on-chip optical interlink. Furthermore, by combining the integration of electrically driven light sources \cite{Liu2019,Paik2019} with h-BN-based waveguide-coupled cavities \cite{Froch2018a,Kim2018,Khelifa2020}, on-chip electrically-pumped lasers can be realized. Due to recent achievements in wafer-scale CVD-grown 2D materials \cite{Lee2018,Chen2020}, the presented prototype platform can be further expanded into larger on-chip photonic circuits with co-integrated optoelectronic devices.\\

\begin{acknowledgments}
This work has been supported by the Swiss National Science Foundation (grant 200020\_192362/1) and the ETH Grant SYNEMA ETH-15 19-1. The authors thank A.~Kuzmina, N.~Fl\"ory, S.~Papadopoulos, A.~Jain, L.~Wang, J.~Huang, Y.~Koyaz, Y.~Xu and U. Koch for fruitful discussions and support. The use of the facilities of the FIRST center for micro- and nanoscience at ETH Zürich is gratefully acknowledged. K.W. and T.T. acknowledge support from the JSPS KAKENHI (Grant Numbers 19H05790, 20H00354 and 21H05233).
\end{acknowledgments}

\section*{Author contributions}
L.N. and R.K. conceived the project. R.K. and S.S. fabricated the devices and performed the experiments. T.T. and K.W. synthesized the h-BN crystals. R.K., S.S. and L.N. analyzed the data and co-wrote the manuscript.

\section*{Supporting Information}
The Supporting Information includes descriptions of: sample fabrication, methods of measurements, details on the efficiency calculations of the LED, PL measurements, characterization of the Gr edge contacts on thick stacks, and details on the determination of the coupling efficiency.\par
\bibliography{references}

\end{document}